\documentclass[epj]{webofc}
\usepackage[varg]{txfonts}   
\usepackage{bm}
\usepackage{slashed}
\usepackage{graphicx}
\usepackage{caption}
\usepackage{subcaption}
\woctitle{MENU 2013}
\begin{document}
\title{Chiral-Scale Perturbation Theory About an Infrared Fixed Point}
%
%

\author{R.~J.~Crewther\inst{1}\fnsep\thanks{\email{rcrewthe@physics.adelaide.edu.au}} \and
        Lewis~C.~Tunstall\inst{1,2}\fnsep\thanks{\email{tunstall@itp.unibe.ch}}\fnsep\thanks{Speaker.}
}

\institute{CSSM and ARC Centre of Excellence for Particle Physics at the Tera-scale, Department of Physics, University of Adelaide, Adelaide SA 5005 Australia
\and
Albert Einstein Centre for Fundamental Physics, Institute for Theoretical Physics, University of Bern, Sidlerstrasse 5, CH-3012 Bern, Switzerland
}

\abstract{%
  We review the failure of lowest order chiral $S\!U(3)_L \times S\!U(3)_R$ perturbation theory $\chi$PT$_3$ to account for amplitudes involving the $f_0(500)$ resonance and $O(m_K)$ extrapolations in momenta. We summarize our proposal to replace $\chi$PT$_3$ with a new effective theory $\chi$PT$_\sigma$ based on a low-energy expansion about an infrared fixed point in 3-flavour QCD. At the fixed point, the quark condensate $\langle\bar{q}q\rangle_\mathrm{vac}\neq 0$ induces nine Nambu-Goldstone bosons: $\pi, K, \eta$ and a QCD dilaton $\sigma$ which we identify with the $f_0(500)$ resonance. We discuss the construction of the $\chi$PT$_\sigma$ Lagrangian and its implications for meson phenomenology at low-energies. Our main results include a simple explanation for the $\Delta I = 1/2$ rule in $K$-decays and an estimate for the Drell-Yan ratio in the infrared limit.
}
\maketitle
\section{Three-flavor chiral expansions: problems in the scalar-isoscalar channel}
\label{intro}
Chiral $S\!U(3)_L \times S\!U(3)_R$ perturbation theory $\chi$PT$_3$ is nowadays well established as the framework to systematically analyze the low-energy interactions of $\pi,K,\eta$ mesons --- the pseudo Nambu-Goldstone (NG) bosons of approximate chiral symmetry.  The method relies on expansions about a NG-symmetry, \textit{viz.}, low-energy scattering amplitudes and matrix elements can be described by an asymptotic series 
\begin{equation}
  {\cal A} = \{ {\cal A}_\mathrm{LO} + {\cal A}_\mathrm{NLO} + {\cal A}_\mathrm{NNLO} + \ldots\}_{\chi\mathrm{PT}_3}
  \label{eqn:series}
\end{equation}
in powers and logarithms of $O(m_K)$ momentum and quark masses $m_{u,d,s} = O(m_K^2)$, with $m_{u,d}/m_s$ held fixed.  The scheme works provided that contributions from the NG sector $\{\pi,K,\eta\}$ dominate those from the non-NG sector $\{\rho,\omega,\ldots\}$; an assumption known as the partial conservation of axial current (PCAC) hypothesis.

\begin{figure}[t]
\centering
\begin{subfigure}[b]{0.52\textwidth}
  \includegraphics[width=\textwidth]{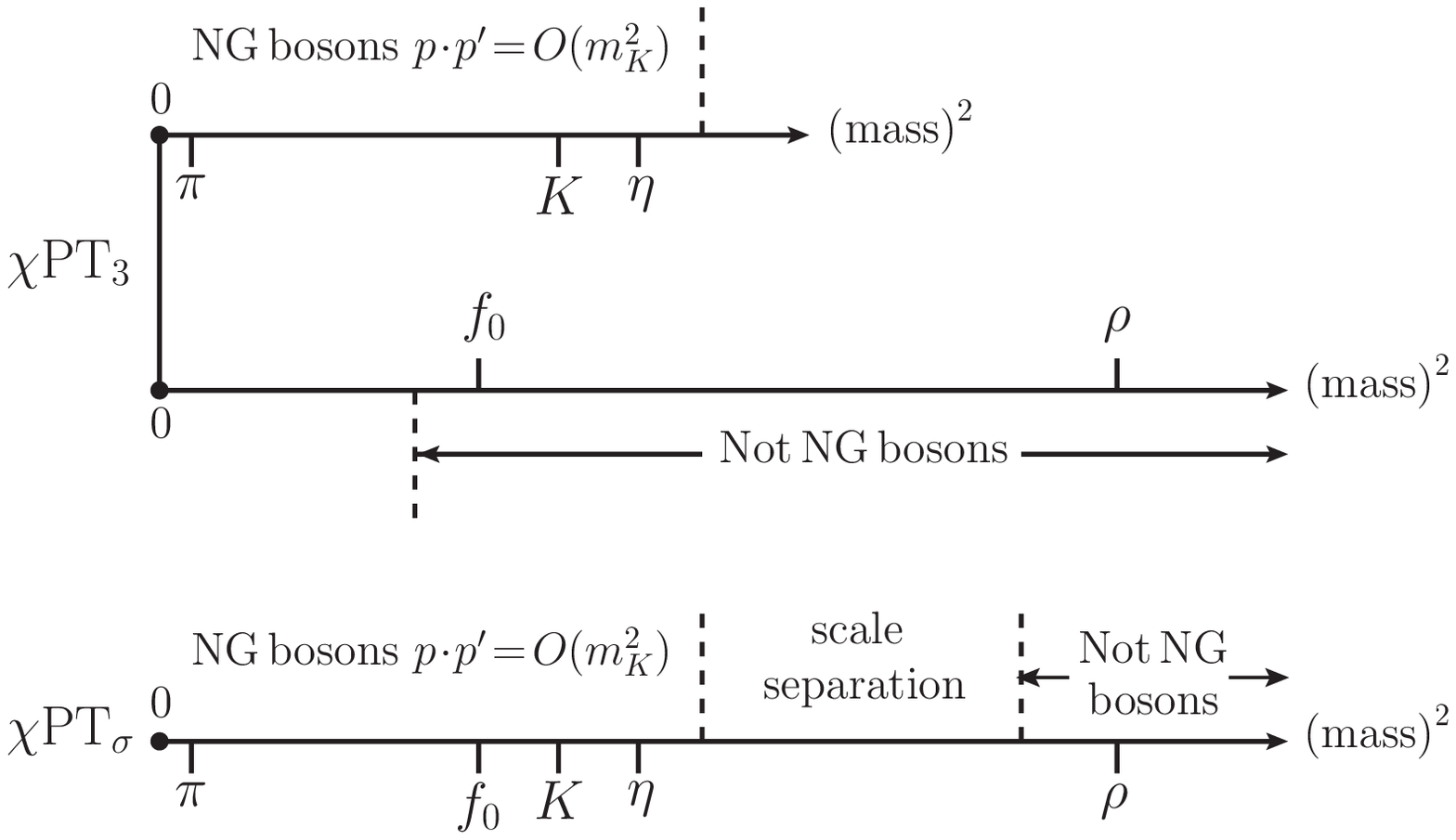}
\caption{Scale separations between Nambu-Goldstone (NG) sectors 
and other hadrons for each type of chiral perturbation theory $\chi$PT 
discussed in this proceeding. 
In conventional three-flavor theory  $\chi$PT$_3$ (top diagram), 
there is {\it no scale separation}: the non-NG boson $f_0(500)$ sits 
in the middle of the NG sector $\{\pi, K, \eta\}$. Our three-flavor
proposal $\chi$PT$_\sigma$ (bottom diagram) for $O(m_K)$ extrapolations
in momenta implies a clear scale separation between the NG sector
$\{\pi,K,\eta,\sigma = f_0\}$ and the non-NG sector $\{\rho,\omega,
K^*,N,\eta',\ldots\}$.}
\label{fig:goldstone}
\end{subfigure}
\quad%
\begin{subfigure}[b]{0.42\textwidth}
\includegraphics[width=\textwidth]{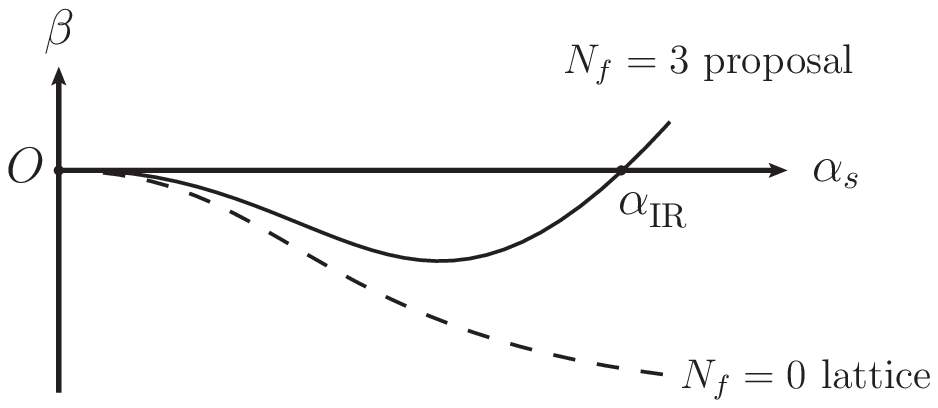}
\caption{Proposed $\beta$-function (solid line) for $N_f = 3$ flavor QCD with infrared fixed point $\alpha_\mathrm{IR}$.  The dashed line shows the Yang-Mills ($N_f=0$) lattice result \cite{Lusc94} for continued growth in $\alpha_s$ with decreasing scale $\mu$.  Despite extensive literature \cite{Alk01} concerning the existence of $\alpha_\mathrm{IR}$, there is currently {\it no consensus} which of the above two, physically distinct, scenarios is actually realized in QCD.  In particular, it is unclear how sensitive existing results are to variations in $N_f$.  This is perhaps unsurprising, since modern calculations utilize different, nonperturbative definitions of $\alpha_s$, thereby making comparisons between various analyses difficult.}
\label{fig:beta} 
\end{subfigure}
\caption{}
\end{figure}

It has been observed \cite{Meiss91}, however, that the $\chi$PT$_3$ expansion (\ref{eqn:series}) is afflicted with a peculiar malady: it typically {\it diverges} for amplitudes which involve both a $0^{++}$ channel and $O(m_K)$ extrapolations in momenta.  
The origin of this phenomenon can be traced to the $f_0(500)$ resonance, a broad $0^{++}$ state whose complex pole mass and residue \cite{Cap06}  
\begin{equation}
m_{f_0} = 441-i\,272 \mbox{ MeV} \quad \mbox{and} \quad |g_{f_0\pi\pi}| = 3.31 \mbox{ GeV}
\label{f_0}\end{equation} 
have been determined to remarkable precision.  
Since $\chi$PT$_3$ classes $f_0$ pole terms as next-to-leading order (NLO), figure~\ref{fig:goldstone} shows why the low-energy expansion (\ref{eqn:series}) fails: the location of $f_0$ and its strong coupling to $\pi,K,\eta$ mesons invalidates the requirements of PCAC.

\section{Three-flavor chiral-scale expansions about an infrared fixed point} 
\label{sec:approximate_scale_invariance}
In this proceeding, we summarize our proposal \cite{us} to solve the convergence problem of $\chi$PT$_3$ expansions (\ref{eqn:series}) by modifying the \textit{leading order} (LO) of the 3-flavor theory.  In short, our solution involves extending the standard NG sector $\{\pi,K,\eta\}$ to include $f_0(500)$ as a QCD dilaton $\sigma$ associated with the {\it spontaneous} breaking of scale invariance.  The scale symmetric counterpart of PCAC --- partial conservation of dilatation current (PCDC) --- then implies that amplitudes with $\sigma/f_0$ pole terms dominate, compared with contributions from the non-NG sector $\{\rho,\omega,K^*,N,\eta',\ldots\}$.%
  \footnote{A discussion on violations of PCDC and Weinberg's power counting scheme \cite{Wei79} in $\gamma\gamma$ channels is contained in \cite{us}.}

This scenario can occur in QCD if at low energy scales $\mu \ll m_{t,b,c}$, the strong coupling $\alpha_s$ for the 3-flavor theory runs \textit{nonperturbatively} to an infrared fixed point $\alpha_\mathrm{IR}$ (figure~\ref{fig:beta}).
At the fixed point, the gluonic term in the strong trace anomaly \cite{Mink76} 
\begin{equation}\label{eqn:anomaly}
\theta^\mu_\mu
 =\frac{\beta(\alpha_{s})}{4\alpha_{s}} G^a_{\mu\nu}G^{a\mu\nu}
         + \bigl(1 + \gamma_{m}(\alpha_{s})\bigr)\sum_{q=u,d,s} m_{q}\bar{q}q
\end{equation}
vanishes, which implies that in the chiral limit
\begin{align}  
\left.\theta^\mu_\mu\right|_{\alpha_s = \alpha_{\mathrm{IR}}}
 = \bigl(1 + \gamma_{m}(\alpha_{\mathrm{IR}})\bigr)
    (m_u\bar{u}u + m_d\bar{d}d + m_s\bar{s}s) \to 0\ , 
\label{scale}\end{align}
and thus $\langle\bar{q}q\rangle_\mathrm{vac}$ acts as a condensate for both scale and chiral $S\!U(3)_L\times S\!U(3)_R$ transformations.%
  \footnote{The former property is a simple consequence of the fact the $\bar{q}q$ is not a singlet under dilatations.  The dual role of $\langle\bar{q}q\rangle_\mathrm{vac}$ was explored \cite{Ell70,RJC70} in some detail prior to the advent of QCD.}
By considering infrared expansions about the combined limit
\begin{equation}
m_{u,d,s} \sim 0 \quad\mbox{and}\quad \alpha_s \lesssim \alpha^{}_\mathrm{IR}  \,,
\label{eqn:exp}
\end{equation}
our proposal is to replace $\chi$PT$_3$ by chiral-scale perturbation theory 
$\chi$PT$_\sigma$, where the strange quark mass $m_s$ in (\ref{scale})
sets the scale of $m^2_{f_0}$ as well as $m^2_K$ and $m^2_\eta$ 
(figure~\ref{fig:goldstone}, bottom diagram). As a result, the rules for 
counting powers of $m_K$ are changed:  $f_0$ pole amplitudes (NLO in  
$\chi$PT$_3$) are promoted to LO. That fixes the LO problem for amplitudes 
involving $0^{++}$ channels and $O(m_K)$ extrapolations in momenta. Note that we achieve this without upsetting successful LO $\chi$PT$_3$ predictions for amplitudes which do not involve the $f_0$; that is because the $\chi$PT$_3$ Lagrangian equals the $\sigma\to 0$ limit of the $\chi$PT$_\sigma$ Lagrangian. 

In the physical region $0<\alpha_s<\alpha_\mathrm{IR}$, the effective theory consists of operators constructed from the $S\!U(3)$ field $U$=$U(\pi,K,\eta)$ and chiral invariant dilaton $\sigma$, with terms classified by their scaling dimension $d$:
\begin{equation} 
\mathcal{L}_{\mbox{\small $\chi$PT$_\sigma$}}
= \mathcal{L}\bigl[\sigma,U,U^\dagger\bigr] =\ :\mathcal{L}^{d=4}_\mathrm{inv} 
 + \mathcal{L}^{d>4}_\mathrm{anom} + \mathcal{L}^{d<4}_\mathrm{mass}:\,.
\label{Lagr}\end{equation}
Explicit formulas for the strong, weak, and electromagnetic interactions are obtained by scaling Lagrangian operators such as $\mathcal{K}\bigl[U,U^\dagger\bigr] 
= \tfrac{1}{4}F_{\pi}^{2}\mathrm{Tr}(\partial_{\mu} U\partial^{\mu}U^{\dagger})$ and $\mathcal{K}_\sigma = \frac{1}{2}\partial_{\mu}\sigma\partial^{\mu}\sigma$ by appropriate powers of the $d=1$ field $e^{\sigma/F_\sigma}$.  For example, the LO strong Lagrangian reads
\begin{align}
\mathcal{L}^{d=4}_\mathrm{inv,\,LO}
 &= \bigl\{c_{1}\mathcal{K} + c_{2}\mathcal{K}_\sigma 
     + c_{3}e^{2\sigma/F_{\sigma}}\bigr\}e^{2\sigma/F_{\sigma}} \,,
\notag \\[1mm]
\mathcal{L}^{d>4}_\mathrm{anom,\,LO} \notag &= \bigl\{(1-c_{1})\mathcal{K} + (1-c_{2})\mathcal{K}_\sigma
      + c_4 e^{2\sigma/F_{\sigma}}\bigr\}e^{(2+\beta')\sigma/F_{\sigma}} \,,
\notag \\[1mm]
\mathcal{L}^{d<4}_\mathrm{mass,\,LO} 
 &= \mathrm{Tr}(MU^{\dagger}+UM^{\dagger})e^{(3-\gamma_{m})\sigma/F_{\sigma}} \,,
\label{Lstr}
\end{align}
where $F_\sigma \approx 100$ MeV is the dilaton decay constant, whose value is estimated by applying an analogue of the Goldberger-Treiman relation to analyses of $NN$-scattering \cite{CC08}.  Here the anomalous dimensions $\gamma_m = \gamma_m(\alpha_\mathrm{IR})$ and $\beta'=\beta(\alpha_\mathrm{IR})$ are evaluated at the fixed point because we expand in $\alpha_s$ about $\alpha_\mathrm{IR}$.  The low-energy constants $c_{1}$ and $c_{2}$ are not fixed by symmetry arguments alone, while vacuum stability in the $\sigma$ direction implies that both $c_3$ and $c_4$ are $O(M)$.  From (\ref{Lstr}), one obtains formulas for the dilaton mass $m_\sigma$
\begin{equation}
m_\sigma^2 F_\sigma^2
= F_\pi^2\bigl(m_K^2 + \tfrac{1}{2}m_\pi^2\bigr)(3 - \gamma_{m})(1 + \gamma_{m})\,,
  - \beta'(4 + \beta')c_4
\label{mass}\end{equation}
and $\sigma\pi\pi$ coupling 
\begin{equation}
\mathcal{L}_{\sigma\pi\pi}
 = \bigl\{\bigl(2+(1-c_1)\beta'\bigr)|\partial \bm{\pi}|^2 
    - (3 - \gamma_{m})m_\pi^2|\bm{\pi}|^2\bigr\}\sigma/(2F_\sigma)\,. \label{Lsigpi}
\end{equation}
Note that (\ref{Lsigpi}) is derivative, so an on-shell dilaton
is $O(m_\sigma^2)$ and consistent with $\sigma$ being the broad resonance 
$f_0(500)$.  

\begin{figure}[t]
\begin{subfigure}[b]{0.6\textwidth}
\center\includegraphics[scale=0.7]{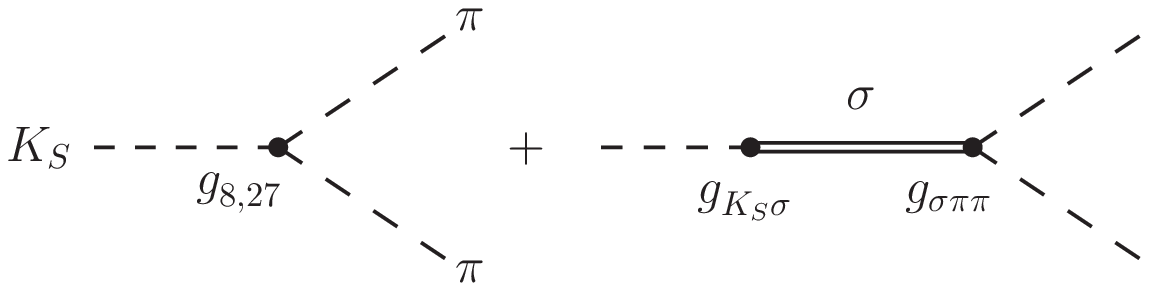}
\caption{Tree diagrams in the effective theory $\chi$PT$_\sigma$ for 
the decay $K_S\to\pi\pi$. The vertex amplitudes due to \textbf{8} and 
\textbf{27} contact couplings $g_8$ and $g_{27}$ are dominated by the 
$\sigma/f_0$-pole amplitude. The magnitude of $g_{K_S\sigma}$ is 
found by applying $\chi$PT$_\sigma$ to $K_S \to \gamma\gamma$ and 
$\gamma\gamma \to \pi\pi$.}
\label{fig:k_pipi}
\end{subfigure}
\qquad%
\begin{subfigure}[b]{0.35\textwidth}
\center\includegraphics[scale=0.65]{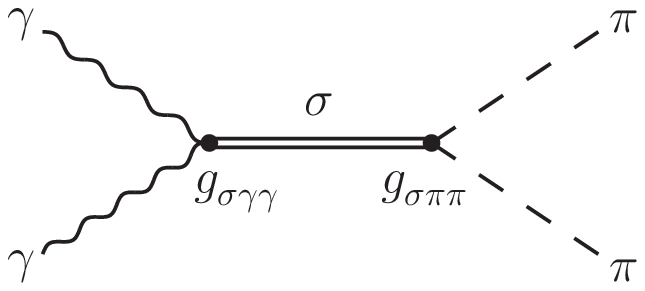}
\caption{Dilaton pole in $\gamma\gamma \to \pi\pi$. Lowest order 
$\chi$PT$_\sigma$ includes other tree diagrams (for $\pi^+\pi^-$ 
production) and also $\pi^\pm,K^\pm$ loop diagrams (suppressed by a factor $1/N_c$) coupled to both 
photons.}
\label{fig:gamgam_pipi}
\end{subfigure}%
\caption{}
\end{figure}

Our proposed replacement for $\chi$PT$_3$ possesses some desirable features, the foremost being:

\begin{enumerate}
  \item The $\Delta I = 1/2$ rule for $K$-decays emerges as a {\it consequence} of $\chi$PT$_\sigma$, with a dilaton pole diagram (figure~\ref{fig:k_pipi}) accounting for the large $I=0$ amplitude in $K_S\to\pi\pi$. Here, vacuum alignment \cite{RJC86} of the effective potential induces an interaction $\mathcal{L}_{K_S\sigma} = g_{K_S\sigma}K_{S}\sigma$ which mixes $K_S$ and $\sigma$ in LO.  The effective coupling $g_{K_S\sigma}$ is fixed by data on $\gamma\gamma\to\pi^0\pi^0$ and $K_S\to\gamma\gamma$, with our estimate $|g_{K_S\sigma}| \approx 4.4\times 10^{3}\,\mathrm{keV}^{2}$
  accurate to a precision $\lesssim 30\%$ expected from a 3-flavor expansion.  Combined with data for the $f_0$ width (Eq.~(\ref{f_0})), we find an amplitude $\left|A_{\sigma\textrm{-pole}}\right| \approx 0.34\,\mathrm{keV}$ which accounts for the large magnitude $|A_{0}|_{\mathrm{expt.}} = 0.33\,\mathrm{keV}$.  Consequently, the LO of $\chi$PT$_\sigma$ explains the 
  $\Delta I = 1/2$ rule for kaon decays.

  \item Our analysis of $\gamma\gamma$ channels and the electromagnetic trace anomaly \cite{RJC72,Ell72} yields a relation between the effective $\sigma\gamma\gamma$ coupling and the nonperturbative Drell-Yan ratio $R_\mathrm{IR}$ at $\alpha_\mathrm{IR}$:
\begin{equation}
g_{\sigma\gamma\gamma} 
=  \frac{2\alpha}{3\pi F_\sigma} \Big(R^{}_{\mathrm{IR}} - \tfrac{1}{2} \Big)\,.
\label{eqn:gsig2gam}
\end{equation}
A phenomenological value for $R_\mathrm{IR}$ is deduced by considering $\gamma\gamma\to\pi^0\pi^0$ in the large-$N_c$ limit (figure~\ref{fig:gamgam_pipi}).  Dispersive analyses \cite{Hof11} of this processes are able to determine the radiative width of $f_0(500)$, which in turn constrains $g_{\sigma\gamma\gamma}$ and yields the estimate $R_\mathrm{IR}\approx 5$.
\end{enumerate}

\end{document}